\documentclass[11pt]{article}
\usepackage{graphicx}
\begin{document}
{\setlength{\oddsidemargin}{1.2in}
\setlength{\evensidemargin}{1.2in} } \baselineskip 0.55cm
\begin{center}
{ { Effect of GUP on Hawking radiation of BTZ black hole}}
\end{center}
\begin{center}
${\rm T.\ Ibungochouba\ Singh^{1},\ Y. \ Kenedy^{2} \ Meitei,\,\,\  and \ I. \ Ablu \ Meitei^{3}}$
\end{center}
\begin{center}
$ ^{1, 2} {\rm Department}$ of Mathematics, Manipur University, Canchipur, 795003, India\\
$ ^{3} {\rm Department}$ of Physics, Modern College, Imphal, Manipur, 795050, India\\
\end{center}
\date{}

\begin{abstract}
 The Hawking radiation of BTZ black hole is investigated based on generalized uncertainty principle effect by using Hamilton-Jacobi method and Dirac equation. The tunneling probability and the Hawking temperature of the spin-1/2 particles of the BTZ black hole are investigated using modified Dirac equation based on the GUP. The modified Hawking temperature for fermion crossing the back hole horizon includes the mass parameter of the black hole, angular momentum, energy and also outgoing mass of the emitted particle.  Besides, considering the effect of GUP into account, the modified Hawking radiation of massless particle from a BTZ black hole is investigated using Damour and Ruffini method, tortoise coordinate transformation and modified Klein-Gordon equation.  The relation between the modified Hawking temperature obtained by using Damour-Ruffini method and the energy of the emitted particle is derived. The original Hawking temperature is also recovered in the absence of quantum gravity effect.
\end{abstract}

1. {\bf Introduction:}

 Hawking [1, 2] discovered the black body radiation of scalar particle based on semiclassical calculation by applying the Wick Rotation method. Refs. [3, 4] showed that the entropy of black hole is proportional to the horizon area of the black hole. Ref [5] proposed the Hawking radiation as a semiclassical tunneling process and their aim is to find well-behaved co-ordinate system at the event horizon to calculate the emission rate. They have shown that Hawking radiation of black hole is not pure thermal if the self gravitation is taken into account. The change of Bekenstein-Hawking entropy was related to the tunneling rate for the Schwarzschild black hole.

The Hawking radiation as a tunneling of extremal and rotating black hole was investigated using Hamilton-Jacobi method and WKB approximation [6] which is an extension of complex path analysis [7]. Refs. [8-10] investigated the tunneling of spin-1/2 and spin-3/2 particle near the black hole horizon applying the WKB approximation. Ref. [11] investigated the tunneling of Dirac particles from a stationary black hole using Pauli Sigma matrices, WKB approximation and Feynman prescription. By choosing appropriate gamma matrices and wave function, the action of radiant particle can be obtained from the Dirac equation which `in turn' is related to the Boltzmann factor of emission according to the semiclassical WKB approximation. Since then, the Hawking radiations of black holes in different space-times were investigated [12-15] using Hamilton-Jacobi method

The presence of minimal length of the order of Plank scale was predicted in different theories of quantum gravity such as string theory, loop quantum gravity, doubly special theory of relativity and Gedanken experiments [16-24]. This observable minimal length leads us to obtain the generalized uncertainty principle(GUP) through modified commutation relation. Refs. [25, 26] derived the expression of GUP as $\Delta x \Delta p \geq \frac{\hbar}{2}[1 + \beta(\Delta p)^{2}]$, where $\beta = \beta_{0}\frac{\ell_{p}^{2}}{\hbar^{2}}$, $\beta_{0}$ is of order unity and $\ell_{p}$ is the  Plank length respectively. The modified commutation relation can be written as
\begin{eqnarray}
[x_{\alpha}, p_{\gamma}] = i\hbar \delta_{\alpha \gamma}[1 + \beta p^{2}],
\end{eqnarray}
 where $x_{\alpha}$ and $p_{\gamma}$ are the modified position and momentum operators respectively, which are also defined by
\begin{eqnarray}
x_{\alpha} &=& x_{0 \alpha}\cr
p_{\gamma} &=& p_{0\gamma}(1 + \beta p^{2}_{0\gamma}),
\end{eqnarray}
 $x_{0\alpha}$ and $p_{0\gamma}$ satisfy the usual commutation relation $[x_{0\alpha},p_{0\gamma}] =i\hbar \delta_{\alpha\gamma}$. It has also been extended the GUP based on doubly special relativity known as DSR-GUP [27, 28]. The Unrugh effect based on modified form of GUP has been investigated in [29]. Refs. [30, 31] discussed the black hole thermodynamics and the tunneling rate was obtained in [32, 33] applying modified from of GUP. Ref. [34] studied creation of scalar particles by an electric field in the presence of quantum gravity effect. Ref. [35] discussed the tunneling of massless scalar particle from Schwarzschild black hole by considering quantum gravity into account influenced by DSR-GUP and Parikh and Wilczek method. Fermion tunneling from Schwarzschild black hole was investigated by applying generalised form of Dirac equation and the resulting remnant of the black hole was discussed in [36]. Quantum gravity effects on the black holes have been discussed in [37-48].

Ref. [49] investigated the Hawking radiation using tortoise co-ordinate transformation in which gravitational field is assumed to be independent of time. The Klein-Gordon scalar field is reduced to a standard form of wave equation near the horizon. By transforming outgoing wave from outside into inside of the horizon, the thermal radiation spectrum of stationary and non-stationary black hole can be obtained. Ref. [50], based on the generalised treatment of barrier penetration proposed by Damour and Ruffini, investigated the tunneling of fermions and bosons crossing black hole horizon and the exact Hawking temperature of Schwarzschild black hole is recovered. Following their work, the Hawking temperature in different space-times have been studied in [51-57].

The aim of this paper is to investigate the tunneling of fermions crossing the horizon of a BTZ black hole by taking quantum gravity effects into account. The correction to Hawking temperature is recovered using generalized Klein-Gordon equation and generalized Dirac equation influenced by GUP.

 The paper is organized as follows. In section 2, the correction of Hawking temperature of BTZ back hole is derived using Dirac equation influenced by GUP. In section 3, utilizing generalized tortoise co-ordinate transformation and Klein-Gordon equation, the Hawking temperture of BTZ black hole has also been discussed with and without the influence of GUP. Some conclusions are given in the last section.

\clearpage

2. {\bf BTZ black hole}

 The line element of BTZ black hole in $(2 + 1)$ dimensional space time is given by [58]
\begin{eqnarray}
ds^2&=& {\Delta}dt^2-\frac{1}{\Delta}dr^2-r^{2} d\phi^2,
\end{eqnarray}
where $\Delta=-M + \frac{r^2}{\ell^2}$. The line element (3) has a singularity at $\Delta = 0$ and the radius of the black hole is given by
\begin{equation}
r_h=\sqrt{M}\ell,
\end{equation}
where $M$ stands for Arnowwitt-Deser-Misner(ADM) mass and expression of ADM is given by
\begin{equation}
M = \frac{{r_h}^2}{\ell^2}.
\end{equation}
The Hawking temperature of BTZ black hole is given by
\begin{equation}
T = \frac{\sqrt{M}}{2 \pi \ell}.
\end{equation}
3. {\bf Tunneling of Dirac particles}

To study the tunneling of Dirac particles from BTZ black hole, the generalized Dirac equation influenced by GUP is given by [36]

\begin{eqnarray}
&& [i\gamma^0\partial_0 + i\gamma^i(1 - \beta{m_0}^2)\partial_i + i\gamma^i\beta\hbar^2(\partial_j\partial^j)\partial_i + \frac{m_0}{\hbar}(1-\beta{m_0}^2 + \beta\hbar^2\partial_j\partial^j)\cr
&& - i\gamma^{\mu}\Gamma_{\mu}(1 + \beta\hbar\partial_j\partial^j - \beta{m_0}^2)]\psi = 0,
\end{eqnarray}
where $\psi$ is a Dirac spinner wave function. For the BTZ black hole in $2 + 1$ dimensional space, $\gamma^{a}$ matrices in $(t, r, \phi)$ coordinate system are chosen as

\begin{eqnarray}
\gamma^t&=& \frac{1}{\sqrt{\Delta}}
\left({\begin{array}{c c}
1 & 0\\
0 & -1\\
\end{array}}\right)\cr
\gamma^r &=& \sqrt{\Delta} \left({\begin{array}{c c}
0 & i\\
i & 0\\
\end{array}}\right)\cr
\gamma^\phi &=& \frac{1}{r}\left({\begin{array}{c c}
0 & 1\\
-1 & 0\\
\end{array}}\right).
\end{eqnarray}

Using he following ansatz for the wave function
\begin{eqnarray}
 \psi(x) = exp(\frac{i}{\hbar}S(t,r,\phi))
 \left(
{\begin{array}{c}
   A(t,r,\phi) \\
   B(t,r,\phi)
 \end{array}}\right),
  \end{eqnarray}
  where $A(t,r,\phi)$, $B(t,r,\phi)$ and $S$ are functions of $t$, $r$ $\phi$ and $S$ is the action of emitted fermion. Ref. [60] showed that the decoupling of Dirac equation could be done only for stationary space time or in the spherically symmetric Vaidya-Bonner black hole [61]. To find the solution of Dirac equation, using Eqs. (3), (8) and (9) in Eq. (7), the two decoupled equations are obtained as
 \begin{eqnarray}
 &&[\frac{1}{\sqrt{\Delta}}(\frac{\partial S}{\partial t}) + m_{0}(1 - \beta m_{0}^{2}) + \beta m_{0}\Delta(\frac{\partial S}{\partial r})^{2} + \frac{\beta m_{0}}{r^{2}}(\frac{\partial S}{\partial \phi})^{2}]A\cr &&+[\frac{(1-\beta m_{0}^{2})}{r}(\frac{\partial S}{\partial \phi}) +\frac{\beta \Delta}{r}(\frac{\partial S}{\partial r})^{2}(\frac{\partial S}{\partial \phi}) + \frac{\beta}{r^{3}}(\frac{\partial S}{\partial \phi})^{3}]B\cr &&+i[(1-\beta m_{0}^{2})\sqrt{\Delta}(\frac{\partial S}{\partial r}) + \beta\Delta\sqrt{\Delta}(\frac{\partial S}{\partial r})^{3} + \frac{\beta \sqrt{\Delta}}{r^{2}}(\frac{\partial S}{\partial r})(\frac{\partial S}{\partial \phi})^{2}]B = 0.
 \end{eqnarray}

 \begin{eqnarray}
&&+[\frac{(1-\beta m_{0}^{2})}{r}(\frac{\partial S}{\partial \phi}) +\frac{\beta \Delta}{r}(\frac{\partial S}{\partial r})^{2}(\frac{\partial S}{\partial \phi}) + \frac{\beta}{r^{3}}(\frac{\partial S}{\partial \phi})^{3}]A\cr
&& - i[(1-\beta m_{0}^{2})\sqrt{\Delta}(\frac{\partial S}{\partial r}) + \beta\Delta\sqrt{\Delta}(\frac{\partial S}{\partial r})^{3} + \frac{\beta \sqrt{\Delta}}{r^{2}}(\frac{\partial S}{\partial r})(\frac{\partial S}{\partial \phi})^{2}]A\cr
&& +[\frac{1}{\sqrt{\Delta}}(\frac{\partial S}{\partial t}) - m_{0}(1 - \beta m_{0}^{2}) - \beta m_{0}\Delta(\frac{\partial S}{\partial r})^{2} - \frac{\beta m_{0}}{r^{2}}(\frac{\partial S}{\partial \phi})^{2}]B = 0
\end{eqnarray}
 The nontrivial solution of the above two equations for $A(t,r,\phi)$ and $B(t,r,\phi)$ will be obtained only when the determinant of the coefficient matrix is zero. Neglecting higher orders of $\beta$, the simplified form of equation is obtained from Eqs. (10) and (11) as
\begin{eqnarray}
&&\frac{1}{\Delta}\left(\frac{\partial S}{\partial t}\right)^2 - \Delta\left(\frac{\partial S}{\partial r}\right)^2 - \frac{1}{r^2}\left(\frac{\partial S}{\partial\phi}\right)^2 - {m_0}^2\cr
&& - \beta\{4\frac{\Delta}{r^2}\left(\frac{\partial S}{\partial\phi}\right)\left(\frac{\partial S}{\partial r}\right)^2 +2\Delta^2\left(\frac{\partial S}{\partial r}\right)^4 + \frac{2}{r^4}\left(\frac{\partial S}{\partial\phi}\right)^4 - 2{m_0}^4\} = 0.
\end{eqnarray}
To investigate the fermions tunneling across the black hole horizon, we need to separate the variables $t,r, \phi$ involved in the above equation.
Taking $S(t, r, \phi) = -\omega t + q\phi + H(r)$, where $\omega$ and $q$ are energy and angular momentum of the particle respectively, and $H(r) = H_{0}(r) + \beta H_{1}(r)$ [59]. The integral of the radial action is obtained as

\begin{eqnarray}
H_{\pm}(r) = \pm\int\frac{\sqrt{\omega^2 -\Delta(\frac{q^2}{r^2}+{m_0}^2)}}{\Delta}[1+\frac{\beta(2{m_0}^2\Delta \omega^2 - \omega^4)}{\Delta\{\omega^2-\Delta(\frac{q^2}{r^2}+{m_0}^2)\}}]dr.
\end{eqnarray}
By taking contour as upper part of semi-circle and using Feynman prescription, the integral of Eq. (13) is computed as

\begin{eqnarray}
H_{\pm}(r) = \pm\frac{\pi i \omega \ell}{2\sqrt{M}}[1+ \beta Y],
\end{eqnarray}
where $H_{+}(r_{h})$ and  $H_{-}(r_{h})$ are outgoing and incoming wave solutions of radial part respectively and $Y = \frac{m_{0}^{2}}{\sqrt{M }\ell}[\frac{2M}{\omega^{2}}(\frac{q^{2}}{M\ell^{2}} + m_{0}^{2}) - 1] - \frac{\omega^{2}}{2M}[\frac{M}{\omega^{2}}(\frac{q^{2}}{M\ell^{2}} + m_{0}^{2}) - 1]$.
The tunneling probability of fermions crossing the black hole horizon is obtained as
\begin{eqnarray}
\Gamma&=& \exp [-\frac{2\pi  \omega \ell}{\sqrt{M}}(1+ \beta Y)].
\end{eqnarray}
The corrected Hawking temperature, $T_D$ of Dirac particle emitted from BTZ black hole after neglecting higher order terms of $\beta$, is obtained as

\begin{eqnarray}
T_D=T(1-\beta Y),
\end{eqnarray}
where $T = \frac{\sqrt{M}}{2 \pi \ell}$ is the Hawking temperature of black hole in the absence of quantum gravity effect. The importance of quantum gravity effect is to lower Hawking temperature of BTZ black hole and the modified Hawking temperature depends on the mass of the black hole and also on the angular momentum, energy and the mass of the emitted Dirac particle. From Eq. (16), we observe the quantum gravity effect prevents the rise of Hawking temperature in BTZ black hole. When $\beta=0$, the standard Hawking temperature of BTZ black hole is recovered.

3. {\bf Tortoise coordinate transformation and tunneling of scalar particles:}

Using tortoise coordinate transformation
\begin{equation}
dr_* = \Delta^{-1}dr,
\end{equation}

the conformally flat two dimensional BTZ black hole is given by
\begin{equation}
ds^2 = dt^2 - dr_*^2.
\end{equation}

From Eq. (2) the square of momentum is given by
\begin{eqnarray}
  p^{\gamma}p_{\gamma}&=&-\hbar^{2}[1-\beta\hbar^{2}(\partial^{\gamma}\partial_{\gamma})]\partial^{\gamma}[1-\beta\hbar^{2}(\partial^{a}\partial_{a})]\partial_{\gamma}\cr
  &\simeq&-\hbar^{2}[\partial^{b}\partial_{b}-2\beta\hbar^{2}(\partial^{b}\partial_{b})\partial^{b}\partial_{b}],
\end{eqnarray}
where the terms of higher order of $\beta$ are neglected.
The Klein-Gordon equation in 1-spatial dimension is defined by $p^{2}\psi(t, x)=(\frac{E^{2}}{c^{2}}-m^{2}c^{2})\psi(t, x)$  with $E=i\hbar\partial_0$ and $p^2=p^{b}p_{b}$.
Then, the two dimensional Klein-Gordon equation of scalar particle can be written as

\begin{eqnarray}
\left[\frac{\partial^2}{\partial t^2} - \frac{\partial^2}{\partial r_*^2} -2\beta\hbar^2\frac{\partial^4}{\partial r_*^4}\right]\psi &=& 0.
\end{eqnarray}
We observe that the space time metric given by Eq. (18) is stationary and for separation of variables in Eq. (20), the wave function $\psi(t,r_*)$
 can be written as
\begin{equation}
  \psi(t,r_*) = e^{-iwt}R(r_*)
\end{equation}
where $w$ denotes the energy of radiating particle. Applying Eq. (21) into Eq. (20), we obtain
\begin{eqnarray}
 2\beta\hbar^2\frac{\partial^4}{\partial r_*^4}R(r_*)+ \frac{d^2}{dr_*^2}R(r_*)+w^2R(r_*)  = 0.
\end{eqnarray}
Then, the following two cases will be discussed:

(1) For the case $\beta = 0:$\\

Eq. (22) becomes the standard wave equation as
\begin{eqnarray}
  \frac{\partial^2}{\partial r_*^2}R(r_*) + w^2R(r_*) &=& 0.\nonumber \\
  \end{eqnarray}

  Here $R(r_*)$ can be written as
  \begin{eqnarray}
   R(r_*)= e^{\eta r_*},
   \end{eqnarray}
  where $\eta$ is given by
  \begin{eqnarray}
  \eta &=& \pm iw.
\end{eqnarray}
The negative solution indicates the incoming wave. Then the outgoing wave solution can be expressed as
\begin{eqnarray}
  \psi(t,r_*) &=& e^{-iwt}e^{iwr_*}
\end{eqnarray}
If we define the advanced time coordinate $v=t+r_*$, the outgoing wave solution can be written as
\begin{eqnarray}
  \psi(v,r_*) &=&  e^{-iwv}e^{2iwr_*}.
\end{eqnarray}
By integrating Eq. (17), the tortoise coordinates transformation is obtained as
\begin{eqnarray}
  r_* &=& \log\left(\frac{r - \sqrt{M}\ell}{r + \sqrt{M}\ell}\right)^{\frac{\ell}{2\sqrt{M}}}.
\end{eqnarray}

The incoming wave has no singularity at the horizon in Eddington coordinate system but outgoing wave has a  logarithmic singularity at the horizon. According to Damour and Ruffini [49] and Sannan [50], the outgoing wave function can continue analytically from outside of the horizon into the inside through the negative half of the complex plane. Therefore, inside the black hole horizon the outgoing wave function can be written as
\begin{eqnarray}
  \psi^{out}(r < r_h) &=& e^{-iwu}\left(\frac{r - \sqrt{M}\ell}{r + \sqrt{M}\ell}\right)^{\frac{iw\ell}{\sqrt{M}}}.
\end{eqnarray}
The outgoing wave outside the horizon is
\begin{eqnarray}
  \psi^{out}(r > r_h) &=&  e^{-iwv}e^{\frac{\pi w\ell}{\sqrt{M}}}\left(\frac{r - \sqrt{M}\ell}{r + \sqrt{M}\ell}\right)^{\frac{iw\ell}{\sqrt{M}}}.
\end{eqnarray}
The tunneling probability across the horizon, $r=r_h$ is
\begin{eqnarray}
  \Gamma &=& \left|\frac{\psi^{out}(r < r_h)}{\psi^{out}(r > r_h)}\right|^2 = e^{\frac{-2\pi w\ell}{\sqrt{M}}}.
\end{eqnarray}
  Thus, the Hawking temperature of a BTZ black hole in the absence of GUP is recovered as
\begin{eqnarray}
 T=\frac{1}{\beta_B}  &=& \frac{\sqrt{M}}{2\pi \ell}.
\end{eqnarray}
(2) For the case of $\beta \neq 0:$

Using Eq. (24) into Eq. (22), we get
\begin{eqnarray}
  \eta^2 &=& \frac{-1\pm\sqrt{1-8\beta\hbar^2w^2}}{4\beta\hbar^2}.
  \end{eqnarray}
  Let us assume $8\beta\hbar^2w^2<<1$ and neglecting higher order terms of $\beta$, we get
  \begin{eqnarray}
  \eta^{2} &=& \frac{-1\pm(1-4\beta\hbar^2w^2)}{4\beta\hbar^2}.
\end{eqnarray}
  There are two values of $\eta$. For positive sign, one value is $\eta=\pm i \omega$  which is the same as $\beta=0$ given in case 1. In such case the original Hawking temperature is recovered and the influence of GUP will also  vanish.

  Here, we are interested in the negative sign because it includes the influence of GUP. Then, the other value of $\eta$ is
\begin{eqnarray}
  \eta &=& \pm iw\sqrt{\frac{1}{2\beta\hbar^2w^2}-1}.
\end{eqnarray}
The above equation has an upper bound for the energy of the particles i.e. $\hbar\omega<\frac{1}{\sqrt{2\beta}}$. In this case there are two values of $\eta$. The positive/negative sign corresponds to outgoing/ ingoing wave. Here, we will discuss only the outgoing wave function as
\begin{eqnarray}
  \psi(t,r_*) &=& e^{-iwt}e^{iwr_*\sqrt{\frac{1}{2\beta\hbar^2w^2}-1}}.
\end{eqnarray}
The generalized tortoise coordinate transformation which is different from [52, 57] is
\begin{eqnarray}
  \hat{r}_* &=& r_*\sqrt{\frac{1}{2\beta\hbar^2w^2}-1}.
\end{eqnarray}
Then the outgoing wave solution can be written
\begin{eqnarray}
  \psi(t,\hat{r}_*) &=& e^{-iwt}e^{iw\hat{r}_*}.
\end{eqnarray}
By defining advanced time coordinate $v= t + \hat{r}_*$, the above equation can be written as
\begin{eqnarray}
  \psi(v,\hat{r}_*) &=& e^{-iwv}e^{2iw\hat{r}_*}.
\end{eqnarray}
Following Damour and Ruffini [49] and Sannan [50], the outgoing wave can be extended inside the horizon as
\begin{eqnarray}
  \psi^{out}(r < r_h)  &=& e^{-iwv}\left(\frac{r - \sqrt{M}\ell}{r + \sqrt{M}\ell}\right)^{\frac{iw\ell}{\sqrt{M}}\sqrt{\frac{1}{2\beta\hbar^2w^2}-1}}.
  \end{eqnarray}
  The outgoing wave outside the horizon $r=r_h$ is
  \begin{eqnarray}
  \psi^{out}(r > r_h) &=& e^{-iwv}\left(\frac{\sqrt{M}\ell - r}{\sqrt{M}\ell + r}\right)^{\frac{iw\ell}{\sqrt{M}}\sqrt{\frac{1}{2\beta\hbar^2w^2}-1}}e^{\frac{\pi w\ell}{\sqrt{M}}\sqrt{\frac{1}{2\beta\hbar^2w^2}-1}}.
  \end{eqnarray}
  The tunneling probability near the horizon  $r=r_h$  at $\beta\neq 0$ is
  \begin{eqnarray}
  \Gamma &=& \left|\frac{\psi^{out}(r<r_h)}{\psi^{out}(r>r_h)}\right|^2 = e^{\frac{-2\pi w\ell}{\sqrt{M}}\sqrt{\frac{1}{2\beta\hbar^2w^2}-1}}.
\end{eqnarray}
From the thermal radiation spectrum of scalar particles, the Boltzmann factor is obtained as
\begin{eqnarray}
  \beta_B &=& \frac{2\pi \ell}{\sqrt{M}}\sqrt{\frac{1}{2\beta\hbar^2w^2}-1}.
\end{eqnarray}
The above expression is the inverse temperature. Assuming $\varepsilon^2 = 2\beta\hbar^2w^2$, the corrected Hawking temperature in Planck scale with the influence of GUP is given by

\begin{eqnarray}
  T &=& \frac{\sqrt{M}\varepsilon}{2\pi \ell}\Big\{1 + \frac{\varepsilon^2}{2} + O(\varepsilon ^{4})\Big\}.
\end{eqnarray}
\begin{figure}
\includegraphics[width=7 cm, height=5 cm]{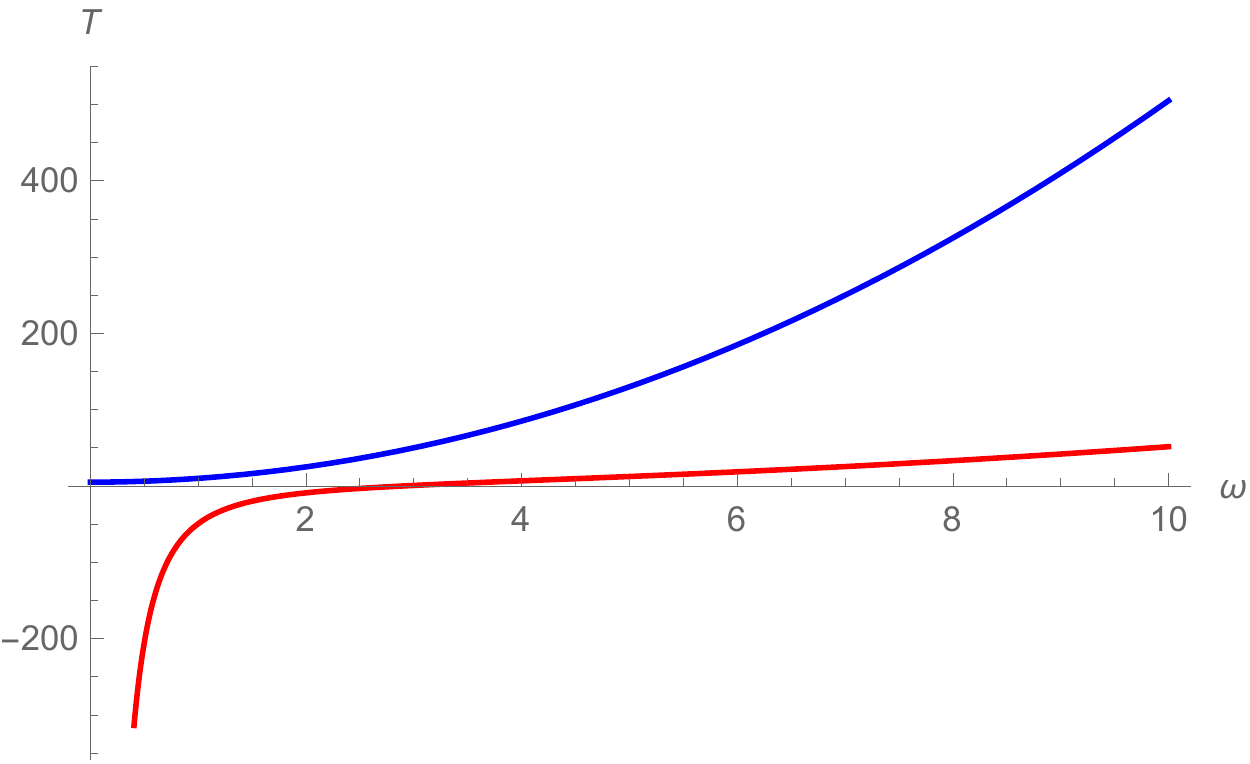}
\caption{Temperature (T) vs energy of the emitted particle ($\omega)$. The blue line corresponds to scalar particles and red line corresponds to Dirac particles. Here, for simplicity, we have taken $\beta=1$, ${\ell=1}, q=1, m_0=1$ and $M=20$. }
\end{figure}
4. {\bf Conclusions and Discussion:}

 In this paper, we have investigated the quantum gravity effect of BTZ black hole by applying the particle tunneling method. To take quantum gravity into account, we take the generalised Dirac equation to discuss the tunneling of fermions crossing the black hole horizon. The tunneling probability of the fermion is calculated and correspondingly the modified Hawking temperature of BTZ black hole is derived. The modified Hawking temperature depends not only mass parameter of the black hole but also angular momentum, energy and the mass of the emitted particles. We also discussed the modified Hawking temperature of BTZ black hole using Damour and Ruffini method, tortoise coordinate transformation and generalized form of Klein-Gordon equation.  In this case, it is shown that the corrected Hawking temperature is related to the energy of the emitted particles.

 From Eq. (14), we observe that
  \begin{itemize}
    \item if $\frac{4m_{0}^{2}\sqrt{M}}{\omega^{2}\ell} > 1 - \frac{\omega^{2}\ell^{2} + 2\sqrt{M}m_{0}^{2}\ell}{q^{2} + M m_{0}^{2}\ell^{2}}$, the corrected Hawking temperature of Dirac particle of BTZ black hole is higher than the standard Hawking temperature.
 \item If $\frac{4m_{0}^{2}\sqrt{M}}{\omega^{2}\ell} = 1 - \frac{\omega^{2}\ell^{2} + 2\sqrt{M}m_{0}^{2}\ell}{q^{2} + M m_{0}^{2}\ell^{2}}$, the GUP effect has been cancelled and the modified Hawking temperature of Dirac particle reduces to the standard Hawking temperature.
 \item  Lastly if $\frac{4m_{0}^{2}\sqrt{M}}{\omega^{2}\ell} < 1 - \frac{\omega^{2}\ell^{2} + 2\sqrt{M}m_{0}^{2}\ell}{q^{2} + M m_{0}^{2}\ell^{2}}$, the modified Hawking temperature of Dirac particle is lower than the standard Hawking temperature.
\end{itemize}
 In the above two cases, the presence of quantum gravity effect prevents the rise of Hawking temperature of BTZ black hole in general. It is observed from Fig. (1) that the Hawking temperature of BTZ black hole for the emission of the Dirac particles can be negative. The spin of the Dirac particles might be responsible for the negative Hawking temperature. Spin systems with bound energy may have negative energy. Negative Hawking temperatures of certain classes of black holes are discussed in [62].

{\bf Acknowledgements} : The YKM acknowledges CSIR (New Delhi) for giving financial support.

\end{document}